\documentclass[twocolumn,preprintnumbers,amsmath,amssymb,pre,groupedaddress,floatfix,showpacs]{revtex4}
\usepackage[dvips]{graphicx}
\newcommand{\omegaf}{\omega_{\rm e}}
\newcommand{\omeganull}{\omega_{\rm 0}}
\begin{document}
\author{Oliver Rudzick}
\email[]{rudzick@fhi-berlin.mpg.de}
\homepage[]{http://www.rudzick.de/}
\affiliation{Abteilung Physikalische Chemie, Fritz-Haber-Institut der Max-Planck-Gesellschaft, Faradayweg 4-6, D-14195 Berlin, Germany}
\title{Oscillating kinks in forced oscillatory media:
A new type of instability}
\begin{abstract}
A new type of instability resulting in oscillatory propagating kinks is
presented. It is observed in periodically forced oscillatory media at 1:1~resonance,
where phase kinks have close similarities to pulses in excitable media.
Considering the periodically forced complex Ginzburg-Landau equation,
examples for transitions involving oscillating kinks between different dynamical regimes are described.
The oscillatory instability is discussed within the framework of a bifurcation analysis of the kink profiles.
\end{abstract}
\maketitle
\section{Introduction}
In many cases, synchronization and desynchronization turned out to be the
underlying mechanisms
for transitions between ordered states and complex behavior in spatially
extended systems~\cite{Pikovsky-Rosenblum-Kurths-Book-01}.
Systems with local oscillatory dynamics, e.g., chemical reactions like
the Belouzov-Zhabotinsky reaction, are sensitive to external
driving~\cite{Kapral-Showalter-Book-95,Petrov-Ouyang-Swinney-97}.
Such systems may be synchronized by external forcing or global feedback.
It has been demonstrated both theoretically and in experiment that these
methods provide powerful tools to control pattern formation in various
systems~\cite{Coullet-Emilsson-92,Coullet-Emilsson-92a,Battogtokh-Mikhailov-96,Battogtokh-Preusser-Mikhailov-97,Kim-Bertram-Pollmann-Oertzen-Mikhailov-Rotermund-Ertl-01,Bertram-Mikhailov-03,Bertram-Beta-Pollmann-Mikhailov-Rotermund-Ertl-03,Bertram-Beta-Rotermund-Ertl-03}.

Detailed studies have been carried out about synchronization phenomena in
1-D geometries~\cite{Chate-Pikovsky-Rudzick-99,Braiman-Ditto-Wiesenfeld-Spano-95,Elphick-Hagberg-Meron-99}.
It turned out that in the presence of external forcing as well
as under global feedback, localized pulses, so-called phase kinks or phase
flips, may appear~\cite{Falcke-Engel-97,Mertens-Imbihl-Mikhailov-93}. 
These objects deserve special interest, since they illustrate the
similarities in the behavior of forced oscillatory media on the one hand and
and excitable systems on the other hand. Coullet first pointed out that 
these phase kinks behave like pulses in excitable
media~\cite{Coullet-Emilsson-92,Coullet-Emilsson-92a}.

These pulses or phase kinks can undergo an instability leading to
self-replication. Then,
spatiotemporal intermittency~\cite{Chate-Manneville-87} can be observed and characteristic
Sierpinski~gasket-looking patterns are possible~\cite{Hayase-Ohta-98}.
Such patterns seem to be
ubiquitous in a wide range of systems.
They have been reported in reaction--diffusion systems~\cite{Lee-McCormick-Pearson-Swinney-94,Lee-Swinney-95},
fluid drag experiments~\cite{Vallete-Jacobs-Gollub-97}, pigmentation dynamics on
sea~shells~\cite{Meinhardt-book-95}, and
dynamics of membrane voltage~\cite{Keener-Sneyd-book-98}.
In~\cite{Argentina-Rudzick-Velarde-04}, a mechanism leading to those
patterns has been described in terms of geometrical arguments.

Here, a new type of instability is presented. It is an oscillatory instability and results in
pulses propagating at constant average velocity with periodically varying shape.  

\section{Model}
In this work, synchronization is studied in a 1-D spatially distributed system close to the onset of an
oscillatory
instability at frequency~$\omeganull$.
A qualitative description of such systems is provided by the complex
Ginzburg-Landau equation~\cite{Aranson-Kramer-02,Cross-Hohenberg-93}
\begin{equation}
        A_t=A-(1+i\alpha)|A|^2 A  +(1+i\beta)A_{xx}.
        \label{eq:CGLE}
\end{equation}
$A(x,t)$ is the complex amplitude of the oscillations.
The parameter $\alpha$ describes the nonlinear frequency shift,
$\beta$ accounts for the dispersion of the medium.
Depending on $\alpha$ and $\beta$, Eq.~(\ref{eq:CGLE}) shows different types
of dynamics.
If $1+\alpha\beta < 0$, two kinds of disordered dynamics are observed:
phase turbulence [$|A|>0$ for all $(x,t)$] and defect turbulence
($|A|=0$ for certain points in the $x$-$t$~plane,
so-called space-time~defects).
In the region  $1+\alpha\beta > 0$, some plane wave solutions are linearly
stable. Under certain conditions, patches of stable plane wave solutions with
different wavelengths coexist separated by localized propagating objects.
Such a state is called spatiotemporal intermittency (STI).

\begin{figure}[htb]
{\includegraphics[width=1.0\columnwidth]{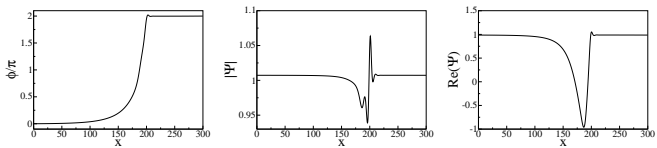}}
\caption{\label{fig:kinkprofile}
Spatial profiles of a propagating phase kink for $\alpha=-0.75$, $\beta=1.8$, $\nu=-0.758$, $B=0.015$.
Right: phase, center: modulus~$|\Psi|$, left: real part $Re(\Psi)$.
}
\end{figure}

\begin{figure}[htb]
\includegraphics[width=\columnwidth]
{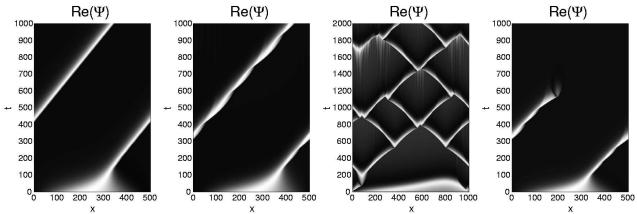}
\caption{\label{fig:kinkbeispiele}
Four typical spatiotemporal diagrams of the real part of $\Psi$.
From left to right: stable propagating kink, oscillatory propagating kink, kink replication, destruction of a kink.}
\end{figure}

In the presence of a spatially homogeneous harmonic forcing, $F(t)=B\exp(i\omegaf t)$,
the above mentioned states may synchronize. This can be modeled with the forced
complex Ginzburg-Landau equation. Using a coordinate frame rotating with the external forcing
[$A\to \Psi\equiv A\exp(-i\omegaf t)$], one obtains an autonomous equation with an additive forcing
term, $F_n=B^n(\Psi^{\ast})^{n-1}$, if the detuning $\nu=\omeganull-m/n\omegaf$ is sufficiently close
to zero~\cite{Coullet-Emilsson-92,Coullet-Emilsson-92a}. Here the 1:1~resonance
($n=m=1$) is considered and the equation reads
\begin{equation}
        \Psi_t=(1+i\nu)\Psi-(1+i\alpha)|\Psi|^2 \Psi  +(1+i\beta)\Psi_{xx}+B\;.
        \label{eq:FCGLE}
\end{equation}
Depending on $B$ and $\nu$, the oscillations in the system can completely synchronize with the external
forcing, resulting in a stationary spatially homogeneous state $\Psi(x,t)\equiv\Psi_0$.
However, this is not the only possible synchronized state.
The fixed point state~$\Psi_0$ can be the background for localized phase jumps of $2\pi$, so-called
phase kinks or phase flips~\cite{Coullet-Emilsson-92,Coullet-Emilsson-92a,Chate-Pikovsky-Rudzick-99}
(Fig.~\ref{fig:kinkprofile}).
They can undergo an instability leading to the destruction of a kink in a space-time defect.
This can give rise to different dynamical regimes~\cite{Chate-Pikovsky-Rudzick-99}
(Fig.~\ref{fig:kinkbeispiele}).

Stable kinks propagating at constant velocity~$V$ are stationary solutions of
\begin{equation}
        \Psi_t=(1+i\nu)\Psi-(1+i\alpha)|\Psi|^2 \Psi  +(1+i\beta)\Psi_{xx}+B-V\Psi_x\,.
        \label{eq:FCGLEv}
\end{equation}
They can be described in a comoving frame $\xi = x-Vt$. Then Eq.~(\ref{eq:FCGLE}) reduces to
a system of four coupled ODEs (traveling wave ODEs). The homogeneous state $\Psi_0$
corresponds to a saddle fixed point with two-dimensional stable and unstable manifolds.
A phase kink is a homoclinic connection between these manifolds. It is of codimension one
and thus exists only for certain values of~$V$. Kink profiles can be computed numerically by
continuing these homoclinic orbits in the traveling wave ODEs, e.g., using AUTO97/2000~\cite{Beyn-Champneys-Doedel-Govaerts-Sandstede-02}.
Such an analysis shows that stable and unstable kink profiles
usually coexist~\cite{Argentina-Rudzick-Velarde-04}.
A linear stability analysis for a kink profile~$\Psi_V$  can be done by inserting~$\Psi_V$ into
the linearized Eq.~(\ref{eq:FCGLEv}) and computing the eigenvalues.

\section{Oscillating propagating kinks}
Numerical simulations indicate that preferably at higher values of
the dispersion coefficient $\beta$, phase kinks can undergo a new type of instability
leading to oscillatory propagating kinks.
Here Eq.~(\ref{eq:FCGLE}) is considered for $\alpha=-0.75$ and
$\beta=1.8$, when the CGLE without forcing exhibits phase turbulence.
In this section, two examples of scenarios leading to oscillating kinks
are discussed. In both cases, $B$ is fixed and the detuning~$\nu$ is varied.
One scenario is connected to the transition from stable kinks to STI,
the other can be observed at higher values of~$B$, when a transition from stable
kinks to a homogeneous state without kinks takes place.

\subsection{Example 1: Oscillating kinks at the transition to STI}
Let us consider the situation at $B=0.015$. For $\nu>\nu_{\rm sn}-0.75858$, only stable
propagating kinks can be observed. If we decrease $\nu$, stable
propagating kinks are still possible. However, depending on the initial
conditions, numerical simulations show a new phenomenon: After a transient,
the spatial profile evolves to a kink profile with a periodically changing
shape propagating at constant average velocity.
For $\nu<\nu_{\rm H}=-0.76081$, any initial condition with a phase rotation of $2\pi$
evolves to such an oscillatory propagating kink; no stable kinks are possible.
At $\nu_{\rm kr}=-0.76598$, a transition to kink replication takes place.
The coexistence of stable and oscillatory kink profiles leads to hysteresis behavior.
Starting with a stable kink at $\nu>\nu_{\rm sn}$, it persists, if $\nu$ is slowly decreased
to $\nu_{\rm H}<\nu<\nu_{\rm sn}$. On the other hand, if one starts with an oscillating kink
at $\nu<\nu_{\rm H}$, increasing of $\nu$ leads to oscillating kinks in the
interval $(\nu_{\rm H},\nu_{\rm sn})$.

\begin{figure}[htb]
\includegraphics[width=1.0\columnwidth]{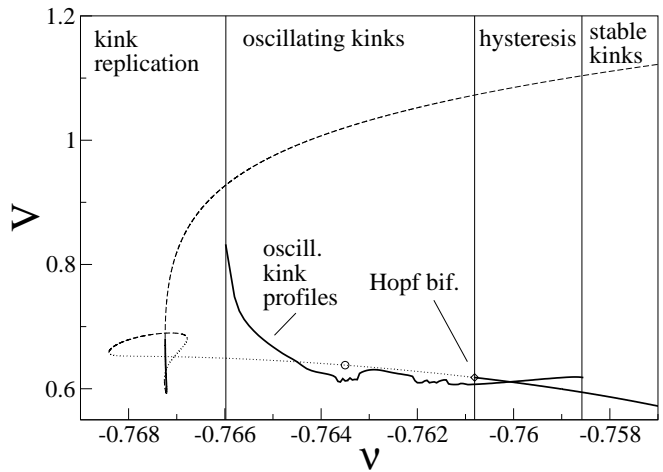}
\caption{\label{fig:osckinks-kba}Bifurcation diagram of the kink profiles at the transition {\em stable kinks} -- {\em oscillating kinks} -- {\em kink replication}
for $\alpha=-0.75$, $\beta=1.8$, $B=0.015$.
Note the complicated bifurcation scenario in the left part of the
diagram. At the intersections of the branches, profiles with different
shapes but same propagation speed coexist.}
\end{figure}

Figure~\ref{fig:osckinks-kba} shows the bifurcations of the kink profiles
for $B=0.015$ with the detuning~$\nu$ as bifurcation parameter.
Solid lines indicate branches of stable attractors, connected to stable kinks
or propagating oscillatory kinks. Dashed and dotted lines indicate branches of unstable
kink profiles (the branch of oscillating kink profiles represents the average velocity
of the profiles).

For $\nu>\nu_{\rm H}$, a linear stability analysis reveals the same situation
as discussed above for the case of stable propagating kinks:
A stable kink profile (solid line in Fig.~\ref{fig:osckinks-kba}) propagating at lower velocity
coexists with a (saddle-)
unstable profile (dashed line in Fig.~\ref{fig:osckinks-kba})  propagating at higher velocity.

\begin{figure}[htb]
\includegraphics[width=1.0\columnwidth]{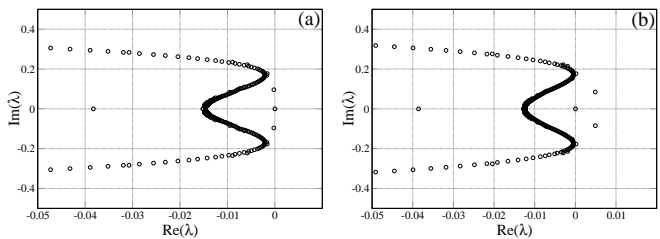}
\caption{\label{fig:osckinks-ew}
(a) Eigenvalues of the linearized operator for $\nu=-0.76081$, $V= 0.61807$.
A pair of complex conjugated eigenvalues crosses the imaginary axis indicating
a Hopf bifurcation. 
(b) Eigenvalues of the linearized operator for $\nu=-0.76350$, $V= 0.63795$
(circle in the diagram Fig.~\ref{fig:osckinks-kba}).
}
\end{figure}

\begin{figure}[htb]
\includegraphics[width=1.0\columnwidth]{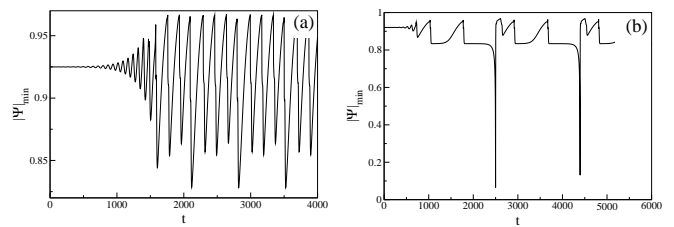}
\caption{\label{fig:osckinks-ts}
(a) Time series of the amplitude minimum in the center of a propagating kink
$\Psi_{\rm min}(t)$,
computed in a comoving reference frame for $\nu=-0.76350$ in the oscillatory
kink regime.
(b) $\Psi_{\rm min}(t)$,
computed in a comoving reference frame for $\nu=-0.76598$, in the kink
replication
regime close to the transition to oscillating kinks.
}
\end{figure}

\begin{figure}[htb]
\includegraphics[width=1.0\columnwidth]{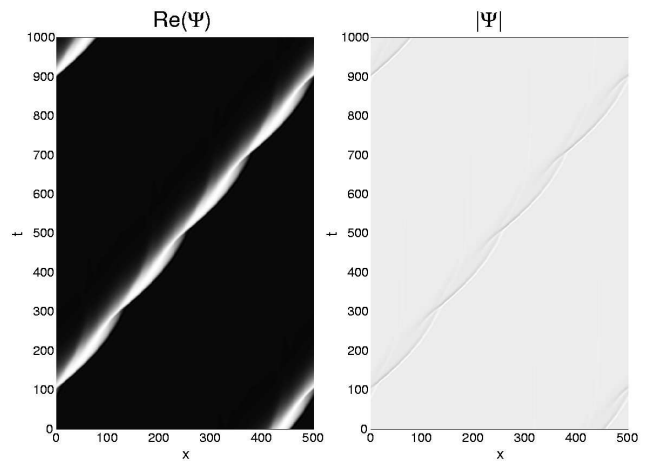}
\caption{\label{fig:osckink-beisp}
Space-time plots of ${\rm Re}(\Psi)$ and $|\Psi|$ for an oscillating kink.
Same parameters as in Fig.~\ref{fig:osckinks-ts}a.
}
\end{figure}

In Fig.~\ref{fig:osckinks-ew}a, the eigenvalues of the linearized operator
are plotted for the stable profile. The curved line of eigenvalues corresponds to the continuous
part of the spectrum. One can see
a pair of complex conjugated eigenvalues just below the real axis.
With decreasing $\nu$, it crosses the real axis. This means that the kink
profile undergoes a Hopf bifurcation changing its stability properties
from those of a stable focus (node) to those of an unstable focus
(node).
For $\nu<\nu_{\rm H}$, the kink profiles corresponding to the lower branch
of the diagram (Fig.~\ref{fig:osckinks-kba}) are unstable with respect to perturbations
leading to oscillatory propagating kinks (for the sake of better readability, such profiles are hereafter
called ``oscillatory unstable'').
In Fig.~\ref{fig:osckinks-ew}b, the eigenvalues of the linearized equation
are presented for such a profile. One sees a pair of complex conjugated
eigenvalues with positive real parts.
The oscillatory instability has been verified in direct simulations of
the CGLE, using a comoving reference frame of a single propagating
kink, with Neumann boundary conditions at the back and Dirichlet
boundary conditions at the front of the kink.
At each time step, the minimum of the modulus of the
field~$|\Psi|_{\rm min}$ in the center of the kink was computed. The time series
 $|\Psi|_{\rm min}(t)$ can be considered as an indicator for the stability
of the kink profile. If the kink is stable (it propagates with constant
shape), $|\Psi|_{\rm min}(t)$ after a short transient reaches a constant value.
For the time series shown in Fig.~\ref{fig:osckinks-ts}a,
the oscillatory unstable kink profile
obtained by numerical continuation with AUTO97 was taken as an initial condition.
$|\Psi|_{\rm min}(t)$ starts oscillating with increasing amplitude and
finally reaches a periodic state.
This indicates that the oscillatory unstable kink profile evolves to a
kink profile with a periodically changing shape propagating
at constant average velocity as shown in Fig.~\ref{fig:osckink-beisp}.
Thus, an oscillatory unstable kink profile coexists with a stable limit cycle
corresponding to propagating oscillatory kinks.

Figure~\ref{fig:osckinks-ts}b illustrates the kink replication process close to
the transition to
oscillating kinks. Initial condition was again the oscillatory
unstable kink profile. $\Psi_{\rm min}(t)$ starts oscillating with increasing
amplitude. At $t>1000$, the oscillation period increases significantly.
 At $t\approx 2500$, $\Psi_{\rm min}(t)$ breaks down to zero periodically,
corresponding to the destruction of the kink and the subsequent creation of a
new pair of counter-propagating kinks.

\begin{figure}[htb]
\includegraphics[width=1.0\columnwidth]{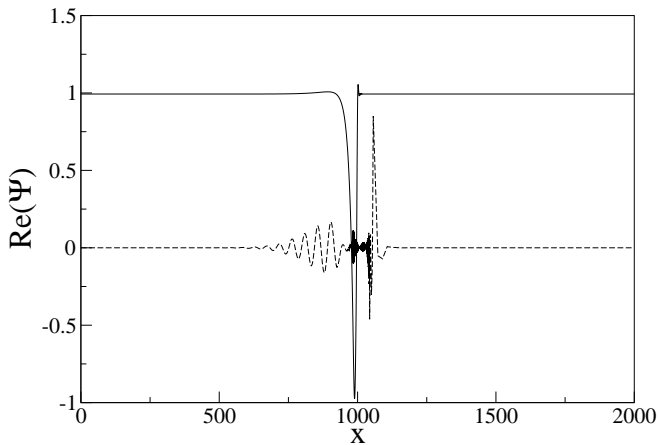}
\caption{\label{fig:osckinks-ev}
${\rm Re}(\Psi)$ for an oscillatory unstable kink profile (solid line) and real part of the
perturbation field acting on ${\rm Re}(\Psi)$ (dashed line). The perturbation field corresponds
to the eigenvector belonging to the unstable pair of complex conjugated eigenvalues of the
spectrum shown in Fig.~\ref{fig:osckinks-ew}b. 
}
\end{figure}

To get more information about the oscillatory instability, the
eigenvectors
belonging to the pair of unstable complex conjugated eigenvalues were computed.
They form a pair of complex conjugated eigenvectors representing the complex amplitude
of the oscillatory perturbation (for the linear stability analysis, the real and
imaginary parts of the complex field~$\Psi$ are treated as independent real variables,
i.e., there are two complex perturbation fields acting on ${\rm Re}(\Psi)$ and
${\rm Im}(\Psi)$, respectively).
In Fig.~\ref{fig:osckinks-ev}, spatial profiles of ${\rm Re}(\Psi)$ and the real part
of the perturbation acting on~${\rm Re}(\Psi)$ are plotted for a spectrum shown in
Fig.~\ref{fig:osckinks-ew}b. One can see an oscillatory perturbation with increasing
amplitude on the tail of the kink. (As the kink is propagating at positive
velocity, ``tail'' refers to the left part of the profile shown in the figure).

\begin{figure}[htb]
\includegraphics[width=1.0\columnwidth]{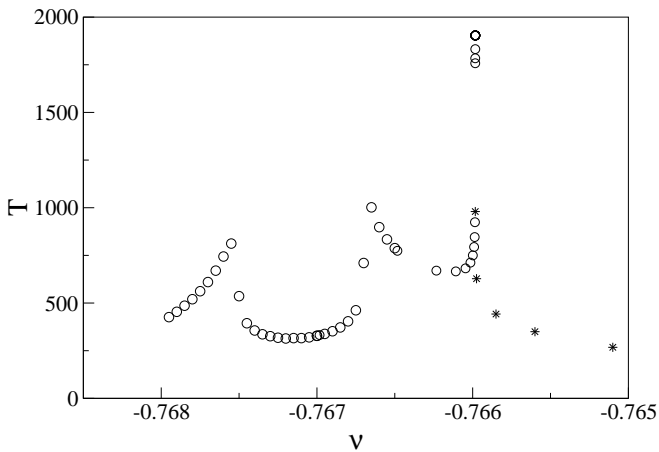}
\caption{\label{fig:tkbr}Period of kink replication~~$T_{\rm kr}$ (circles) and kink
oscillation period~$T_{\rm o}$ (stars) vs.~$\nu$ at the transition {\em oscillating kinks} -- {\em kink replication} 
(parameters: $\alpha=-0.75$, $\beta=1.8$, $B=0.015$).}
\end{figure}

For further decreasing~$\nu$, one can observe that the oscillation
period~$T_{\rm o}$
of the propagating kink profiles increases.
At $\nu=\nu_{\rm kr}$, $T_{\rm o}$ diverges and the kink profile evolved from the
oscillatory unstable profile
remains for a long time close to the coexisting saddle profile.
If $\nu<\nu_{\rm kr}$, 
the kink profiles are destroyed and the kink replication process
begins.
Figure~\ref{fig:osckinks-ts}b shows $|\Psi|_{\rm min}(t)$ in the kink
replication
regime close to the transition to oscillating kinks. If a kink is destroyed
in a defect,  $|\Psi|_{\rm min}(t)$ drops to zero.
The periods of the kink oscillations~$T_{\rm o}$ and the kink replication~$T_{\rm kr}$
are plotted vs. $\nu$ in Fig.~\ref{fig:tkbr}. The divergence of both $T_{\rm o}$
and~$T_{\rm kr}$ at $\nu\approx\nu_{\rm kr}$ can be clearly seen.

\begin{figure}[htb]
\includegraphics[width=1.0\columnwidth]{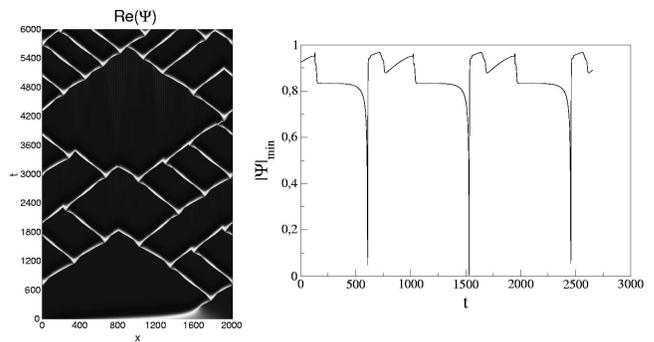}
\caption{\label{fig:osckink-sti}Kink replication involving oscillating kinks at $\alpha=-0.75$, $\beta=1.8$, $B=0.015$, $\nu=-0.765985$.
{\em Left:} space-time plot of the real part of the complex oscillation amplitude~${\rm Re}(A)$. {\em Right:}
$|\Psi|_{\rm min}(t)$ computed in a comoving reference frame as in Fig.~\ref{fig:osckinks-ts}. 
}
\end{figure}

The kink replication process leads to characteristic Sierpinski~gasket-like spatiotemporal pattern which can be
observed in a wide range of systems. The resulting dynamics can be seen as a form of~STI~\cite{Chate-Pikovsky-Rudzick-99}.
Close to the transition at $\nu=\nu_{\rm kr}$, a new form of STI involving oscillating kinks can be found.
In the left panel of Fig.~\ref{fig:osckink-sti}, a space-time plot of such STI is shown. The time series $|\Psi|_{\rm min}(t)$
in the right panel of Fig.~\ref{fig:osckink-sti} illustrates the oscillatory transients of a kink between two
subsequent
kink replication events.

Although the kink replication period~$T_{\rm kr}$ obviously diverges, if $\nu$ approaches $\nu_{\rm kr}$,
it was not possible
to find a scaling law. The appearance of additional local maxima of $T_{\rm kr}(\nu)$ for $\nu<\nu_{\rm kr}$ in Fig.~\ref{fig:tkbr}
indicates that the parameter dependence of $T_{\rm kr}$ is very complicated. Thus, simple scaling
laws might fit only in small parameter intervals which cannot be resolved numerically. 

The numerical results can be interpreted as follows:
The transition from stable kinks to a hysteresis regime, where stable kinks
and oscillating kinks coexist as attractors with decreasing $\nu$ is typical
for a saddle-node bifurcation of limit cycles. Branches of stable and unstable
limit cycles are created. The unstable limit cycle corresponds to a localized oscillatory
propagating object that acts as a separatrix in the phase space.
It separates spatial profiles that evolve to a stable kink from
those evolving to an oscillating kink. Another indicator for a saddle-node
bifurcation of limit cycles is the fact that the oscillations of the kink
profile have a finite amplitude at their onset.

With decreasing detuning $\nu$, the stable kink profiles associated with
the lower branch in the bifurcation diagram (Fig.~\ref{fig:osckinks-kba})
undergo a Hopf~bifurcation. This Hopf~bifurcation is subcritical:
The unstable oscillatory kink profile merges with the stationary kink profile
associated with that branch. As a consequence, they undergo an oscillatory
instability leading to the coexistence of an oscillatory unstable kink profile
and a stable
limit cycle.
If $\nu$ is further decreased, the stable limit cycle collides with the
saddle kink profile corresponding to the upper branch in the bifurcation
diagram (Fig.~\ref{fig:osckinks-kba}) and is destroyed in a homoclinic
Andronov~bifurcation. The divergence of the kink oscillation period is an
indication for this type of bifurcation. Numerical simulations at $\nu\approx\nu_{\rm kr}$
reveal that the oscillating kink profiles remain for a long time close to the saddle profile
corresponding to the upper branch in the bifurcation diagram (Fig.~\ref{fig:osckinks-kba})
\footnote{Animations showing the temporal evolution of kink profiles in a comoving reference
frame can be found at {\tt http://www.rudzick.de/kink.html}.}.
Consequently, the average velocity of the oscillating kink profiles converges to
the velocity of the saddle kink profiles for $\nu\to\nu_{\rm kr}$~\footnote{This may happen in a very narrow
parameter interval which cannot be resolved numerically. Thus, the branch corresponding to oscillating propagating
kinks in the diagram (Fig.~\ref{fig:osckinks-kba}) ends before reaching the saddle kink profile branch.}.

At $\nu>\nu_{\rm kr}$, no stable attractor associated to propagating kinks exists, and any
kink-like object will be destroyed in a defect after a certain transient.
In the above discussed case with $B=0.015$ this leads to kink replication.

\subsection{Example 2: Transition stable kinks -- no kinks}
\begin{figure}[htb]
\includegraphics[width=1.0\columnwidth]{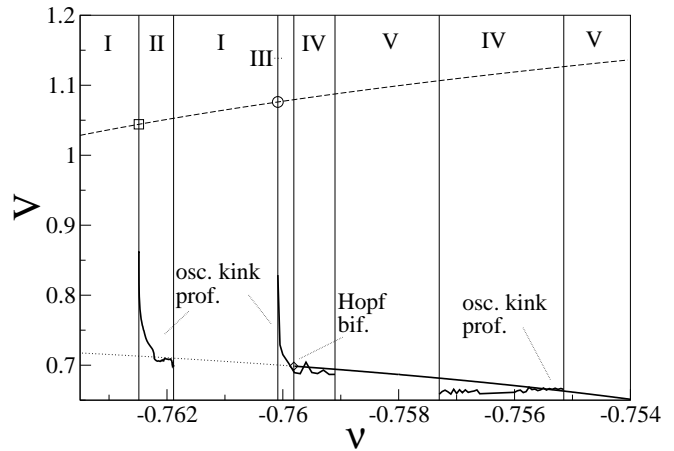}
\caption{\label{fig:osckinks-nk}
Bifurcations of the kink profiles ($B=0.022$, other parameters as in
Fig.~\ref{fig:osckinks-kba}).
Dashed (dotted) lines denote branches of saddle (oscillatory unstable/unstable node) kink profiles.
The Roman numerals denote the different dynamical regimes: (I) {\em No kinks},
(II) bistability {\em no kinks} -- {\em oscillating kinks}, (III) {\em oscillating kinks}, (IV) multistability {\em oscillating kinks} --
{\em stable kinks}.
}
\end{figure}

More complicated scenarios can be encountered at the transition {\em stable
kinks} -- {\em oscillatory propagating kinks} -- {\em no kinks}
for higher values of the forcing amplitude $B$. 
In Fig.~\ref{fig:osckinks-nk}, a bifurcation diagram for $B=0.022$ is
presented, again with $\nu$ as the bifurcation parameter. As in the previously
discussed case for $B=0.015$, the upper branch in the diagram corresponds
to saddle unstable kink profiles, since the lower branch is associated
to oscillatory unstable kink profiles resulting from a Hopf bifurcation
of the stable profiles, which are indicated by a solid line.
Oscillatory propagating kink profiles are again indicated by their average velocity.
Starting decreasing $\nu$ in the stable kink regime at the right boundary of the diagram in Fig.~\ref{fig:osckinks-nk}, one first encounters a region with multistability for
$-0.75515 > \nu > -0.7573$ (IV in Fig.~\ref{fig:osckinks-nk}).
In this parameter interval, two stable attractors coexist corresponding to
stable and oscillating kinks. It is separated by an intermediate stable kink region (V in Fig.~\ref{fig:osckinks-nk}) from another multistable region at lower values of $\nu$. Therefore,
oscillatory kinks in this multistable region can only be found by starting in an oscillatory
kink region and varying both $B$ and $\nu$.
At $\nu=-0.7591$, a branch of oscillating kink profiles
appears, leading again to multistability and hysteresis between stable and oscillatory kinks
(IV in Fig.~\ref{fig:osckinks-nk}).
At $\nu=-0.75981$, the branch of stable kink profiles is turned into a branch of oscillatory
unstable profiles in a Hopf~bifurcation, and only oscillatory kinks are possible
(III in Fig.~\ref{fig:osckinks-nk}). At $\nu_0=-7600868$, a transition
{\em oscillating kinks} -- {\em no kinks} takes place. However, in contrast to the situation at the
transition
to the {\em kink replication} regime, at $\nu_1=-0.7618857$, a new branch of oscillatory kink profiles
appears.
In this regime (II in Fig.~\ref{fig:osckinks-nk}), oscillatory kinks can be found only
for certain initial conditions with a phase rotation of~$2\pi$, whereas in the oscillatory kink regime at
higher values of $\nu$ (III in Fig.~\ref{fig:osckinks-nk}) each such initial condition
$2\pi$ seems to evolve to a kink profile.
This bistable region disappears at $\nu_2=-0.7624856$ in another transition from {\em oscillating kinks}
to a
homogeneous state without kinks (I in Fig.~\ref{fig:osckinks-nk}).

Let us now have a closer look at the transitions to a spatially homogeneous state without
kinks. If those transitions occur with decreasing $\nu$ (at $\nu=\nu_0$ and $\nu=\nu_2$),
the kink oscillation period was found to increase, when the transition was approached.

\begin{figure}[htb]
\includegraphics[width=1.0\columnwidth]{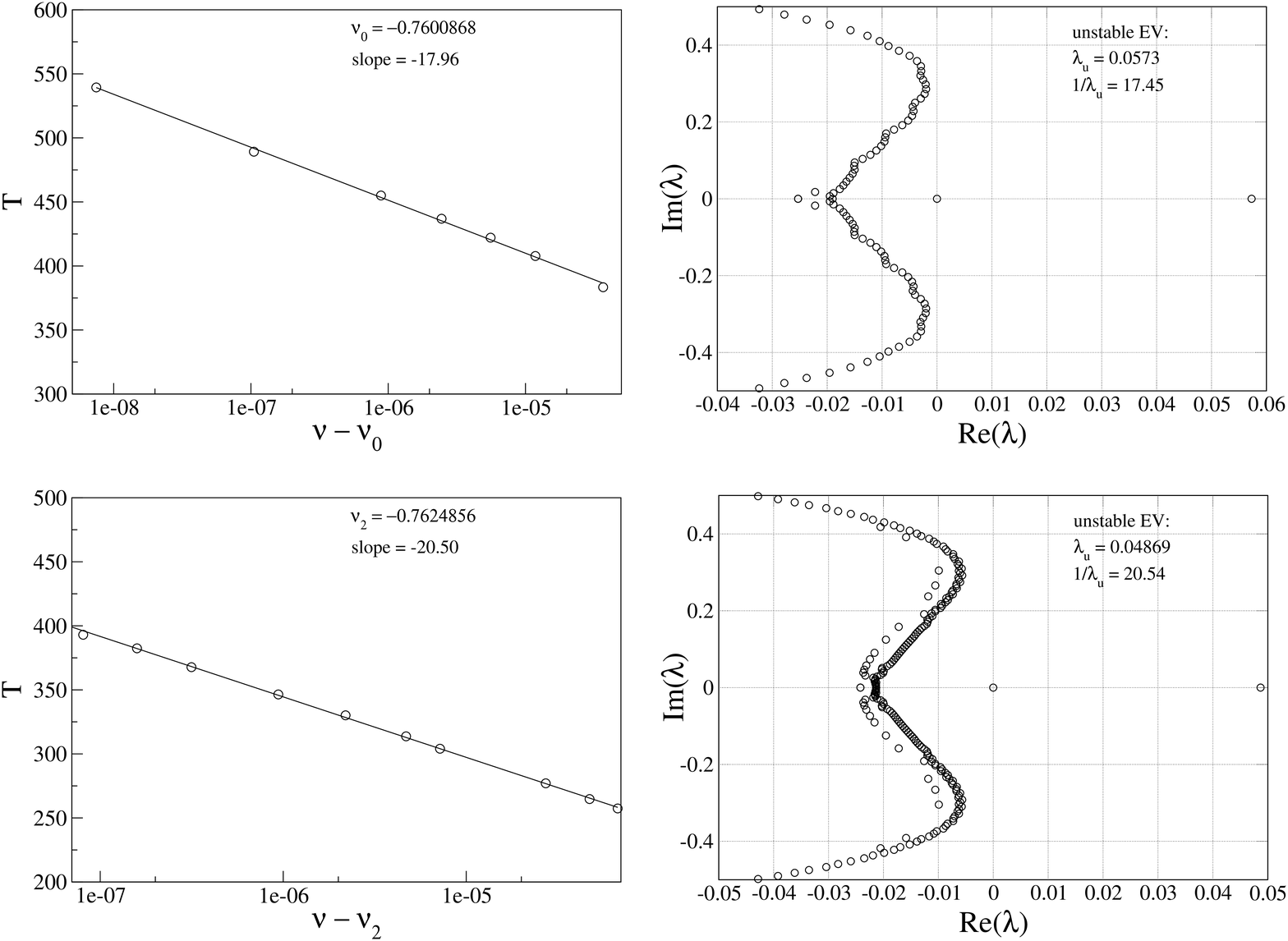}
\caption{\label{fig:osckinks-nk-logdiv}
{\em Upper panel}
Left: Semi-log plot of the kink oscillation period $T_{\rm osc}$ vs. distance to
the transition $\nu-\nu_0$.
Right: Eigenvalues of the linearized operator for $B=0.022$, $\nu=-0.7600868$, $V=1.07608$ (circle in Fig.~\ref{fig:osckinks-nk}).
A positive real eigenvalue $\lambda=0.05373$ indicates a saddle.\\
{\em Lower panel}
Left: Semi-log plot of the kink oscillation period $T_{\rm osc}$ vs. distance to
the transition $\nu-\nu_2$.
Right: Eigenvalues of the linearized operator for $B=0.022$, $\nu=-0.7624856$, $V=1.04418$
(square in Fig.~\ref{fig:osckinks-nk}).
}
\end{figure}

To get a further insight into the nature of the transition, the oscillation period $T_{\rm osc}$ of the
oscillating kinks was measured from the time series $|\Psi|_{\rm min}(t)$ for $\nu>\nu_0$ and $\nu>\nu_2$.
The results are presented in the left panel of Fig.~\ref{fig:osckinks-nk-logdiv} as
semi-log plots vs. distance to the transition $\nu-\nu_0$ and $\nu-\nu_2$, respectively.
The oscillation period diverges in both cases as $T_{\rm osc}=1/\lambda\log(\nu-\nu_0)$
[$T_{\rm osc}=1/\lambda\log(\nu-\nu_2)$]. 
The right panel of Fig.~\ref{fig:osckinks-nk-logdiv} shows the result of the linear stability
analysis performed for kink profiles belonging to the upper branch in the
bifurcation diagram at  $\nu=\nu_0$ and $\nu=\nu_2$, respectively.
One can see one real eigenvalue beyond the imaginary axis indicating a saddle.
The values of~$\lambda$ obtained by fitting the logarithmic
divergence law with the unstable eigenvalue of the
linearized operator are in good agreement.
This is again a clear indication of a homoclinic Andronov bifurcation involving the branch of
saddle unstable kink profiles shown as dashed line in Fig.~\ref{fig:osckinks-nk}.

At $\nu=\nu_1$, where the transition to a homogeneous state without kinks takes place, if $\nu$
is increased, no divergence of the kink oscillation period was found. However, at $\nu>\nu_1$,
oscillating kink profiles exist as transients before being destroyed via the creation of a defect.
The length of these transients increases when approaching the transition. This can be seen
as an indicator for a saddle-node bifurcation of the oscillatory kink profiles.
The appearance of intermediate regimes with oscillatory kinks might be connected
to a complicated bifurcation scenario of limit cycles associated with propagating oscillating
kinks, similar to that observed for stationary kink profiles.  

\section{A phase diagram}
\begin{figure}[htb]
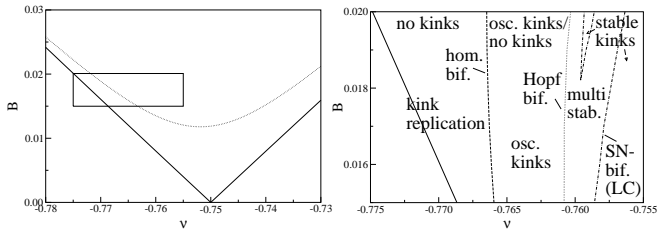

\includegraphics[width=0.5\columnwidth]{figure12l.eps}\includegraphics[width=0.5\columnwidth]{figure12r.eps}
\caption{\label{fig:phasediagr}
Rough phase diagram in the $\nu$-$B$~plane.
{\em Left:} Boundary of the region, where spatially
homogeneous steady states exist
(Arnold tongue, solid line), and stability boundary
of the spatially homogeneous state (dotted line). Below the line, the spatially homogeneous state
becomes linearly unstable.
{\em Right:}
Enlargement of the rectangular-shaped region in the left panel.
The solid line denotes again the boundary of the
Arnold tongue. Additionally, the lines, where the bifurcations discussed in the text take place,
are shown.
}
\end{figure}

Studies of transitions to the regime with stable propagating
kinks have been carried out for a wide range of parameters.
A rough phase diagram showing the different regimes in the $B$-$\nu$~plane
is given in Fig.~\ref{fig:phasediagr}.

The dashed line in Fig.~\ref{fig:phasediagr} indicates homoclinic bifurcations at the
transition {\em no kinks} -- {\em oscillating kinks} and {\em kink replication} -- {\em oscillating kinks}
(the line corresponding to the transition between {\em kink replication} and {\em no kinks} is not shown).
At higher values of $B$, several transitions {\em no kinks} -- {\em oscillating kinks} are
possible, when $\nu$ is varied.
 The dashed line in Fig.~\ref{fig:phasediagr} always corresponds
to the last appearance of oscillating kinks, when $\nu$ is decreased.
This means that on the left side of the dashed line in Fig.~\ref{fig:phasediagr}
no oscillatory kinks are possible, whereas on the right side of this line oscillating kinks
can be found except for some small parameter regions in the upper part of the diagram at higher
values of~$B$.
The Hopf bifurcation of kink profiles is shown as dotted line.
It separates the region with oscillatory kinks from the bistable region, where
oscillatory and stable kinks coexist.
The dashed-dotted line marks the saddle-node bifurcation of oscillatory kink profiles.
It is the boundary between the bistable region and the regions, where only stable kinks
exists. The appearance of intermediate regions with stable kinks or bistability again indicates
complicated bifurcation scenarios of limit cycles corresponding to oscillatory kinks
at higher values of $B$

The locations of the bifurcation lines in the parameter space show that
the scenarios presented above are typical
for transitions from stable kinks to {\em kink replication} or to {\em no kink} regimes,
reflecting the tendency to more complicated
transition scenarios at higher values of $B$.

\section{Conclusions}
The transitions {\em stable kinks} -- {\em kink replication} and {\em stable kinks} -- {\em no kinks}
studied in this work are connected to the synchronization of phase turbulence in the CGLE by external forcing.

It turned out that
the parameter regions, where kink profiles exist,
are not indentical to those with stable kinks.
The boundaries of the latter are determined by instabilities of the kink profiles. This leads to
intermediate regimes with new types of attractors such as limit cycles
associated with oscillatory propagating kinks. Multistable regimes
where stable kinks and oscillating kinks as attractors coexist are also possible.
The destruction of those
attractors is responsible for the transition either to {\em kink replication} or
to a spatially homogeneous state.
Instead of a single saddle-node bifurcation, the kink profiles undergo
complicated bifurcation scenarios leading to the coexistence of
several kink profiles. In those scenarios the creation of branches of
stable kinks was also observed. As a consequence, transition scenarios
with intermediate regimes of stable kinks are possible.

Moreover, at the transition from oscillatory kinks to a spatially homogeneous state
without kinks intermediate regions with oscillatory kinks can be encountered.
A possible explanation could be that their existence is due to a complicated bifurcation scenario
of the limit cycles corresponding to oscillating propagating kinks involving saddle-node
bifurcations of limit cycles. Another indication for
this assumption is the appearance of an intermediate stable kink region in the phase diagram
in Fig~\ref{fig:phasediagr}.
In the present work, these saddle-node bifurcations of limit cycles were only observed in connection
with transitions {\em oscillating kinks} -- {\em stable kinks} or {\em oscillating kinks} -- {\em no kinks}. However,
if one varies the values of $\alpha$ and $\beta$, saddle-node bifurcations connected to transitions
{\em oscillating kinks} -- STI might be possible. Perhaps, such a transition could lead to new forms
of STI. 

The oscillating propagating phase kinks are the result of a new type of instability
which is presumably favored at higher values of the dispersion coefficient $\beta$ of the CGLE.
Numerical studies have been carried out revealing very complicated bifurcation scenarios of the
kinks involving non-coherent structures like oscillating kinks. These scenarios are not yet
completely understood. One can assume that they occur in a wide range of parameters.
The present work can be a starting point for further systematical investigations.

\begin{acknowledgments}
The author thanks A.~S.~Mikhailov for valuable discussions and critical reading of the manuscript,
M.~Argentina, H.~Chat{\'e}, A.~S.~Pikovsky, M.~G.~Velarde for stimulating discussions and I.~Reinhardt
for proofreading.
\end{acknowledgments}

\newpage
\bibliographystyle{apsrev}
\bibliography{rud,pap-ab,pap-ce,pap-fg,pap-hj,pap-kl,pap-mn,pap-oq,pap-rs,pap-tz,books}

\end{document}